# Antisymmetric chirp transfer to high-energy ultraviolet pulses via gas-based chirped four-wave mixing

LINSHAN SUN,[1,2*] HAO ZHANG,[1,3] CAMERON LEARY,[1] CONNOR LIM,[1] CHAD PENNINGTON,[1] GIA AZCOITIA,[1] BRITTANY LU,[1] AND SERGIO CARBAJO[1,2,3,4]

[1]*Department of Electrical & Computer Engineering, University of California Los Angeles, Los Angeles, CA 90095, USA*
[2]*California NanoSystems Institute, 570 Westwood Plaza, Los Angeles, CA 90095, USA*
[3]*SLAC National Accelerator Laboratory, Stanford University, Menlo Park, California 94025, USA*
[4]*Physics and Astronomy Department, University of California Los Angeles, CA 90095, USA*
*lssun@ucla.edu

**Spectro-temporal shaping of high-power femtosecond ultraviolet (UV) pulses remains a key challenge in ultrafast optics. Tailoring high-energy, ultrashort UV pulses underpins applications in ultrafast dynamics, high-precision spectroscopy, strong-field physics, charged-particle radiation sources, and industrial microfabrication. However, the transmission and damage threshold of the dielectric optics limit programmable shaping for high-power UV pulses. Towards overcoming this challenge, we demonstrate an antisymmetric dispersion transfer from near-infrared (NIR) pulses to UV pulses via chirped four-wave mixing (CFWM) in noble gas, whereby negatively chirped NIR pulses map quasi-linearly to positively chirped UV in a gas-filled hollow capillary fiber (HCF) at the expense of a moderate conversion efficiency of 13%. This antisymmetric chirp transfer approach broadens the basis for tailoring UV pulses by leveraging the high damage tolerance, large bandwidth acceptance, and intrinsically low dispersion of noble gases, rather than relying on conventional nonlinear crystals.**

**Introduction.** Ultrafast optics has become a cornerstone of modern science and technology[1], providing a powerful platform for understanding and controlling light-matter interactions. Within this context, ultrashort pulse shaping[2] that controls a pulse's temporal, spectral, and phase characteristics has emerged as an essential technique. However, conventional high-power pulse shapers are typically limited to wavelengths above the visible range due to the restricted transmission window of dielectric materials[3]. Meanwhile, the tunable UV spectrum and temporal profile[4, 5] are crucial in ultrafast optics, as UV photons can resonantly interact with excited states[6], coherently control molecular dynamics[7], enable molecular fingerprinting[8] and photoemission[9–11] on attosecond-to-femtosecond timescales. For instance, shaped UV femtosecond pulses control localized transient electron dynamics during photon-matter interactions, since the subsequent energy transfer from electrons to ions is of picosecond order, thus lattice motion is negligible within the femtosecond pulse duration[6]. In particular, negatively chirped UV pulses have broad applicability for pulse compression[12], ultrafast spectroscopy[13], and driving high harmonic generation toward ultrafast X-ray generation[14].

Ultrashort UV pulses are commonly generated by sum-frequency generation (SFG) in nonlinear crystals. While efficient, crystal-based approaches suffer from pulse broadening due to group-velocity walk-off[15] and limited phase-matching bandwidth arising from the strong material dispersion. Direct UV pulse shaping, such as using liquid-crystal spatial light modulators, acousto-optic modulators, and acousto-optic programmable dispersive filters, typically faces constraints in shorter wavelengths, bandwidth, or average power[16, 17]. In contrast, CFWM nonlinear processes[12, 18] can effectively mitigate issues such as severe temporal walk-off[19], low damage threshold, and significant energy loss. HCF can handle millijoule, few-cycle pulses[20] and up to 318 $W$ average power[21], as well as providing nonlinearity to transfer the spectrum and phase between different frequencies. It also provides octave-spanning phase-matching bandwidth compared with bulk media, while dramatically enhancing the light-matter interaction as compared with gas-based techniques in free space[22]. Motivated by these advantages, we aim to generate broadband, spectrally shaped UV light in gas-filled systems. In such configurations, chirp control can be simultaneously achieved through this indirect method.

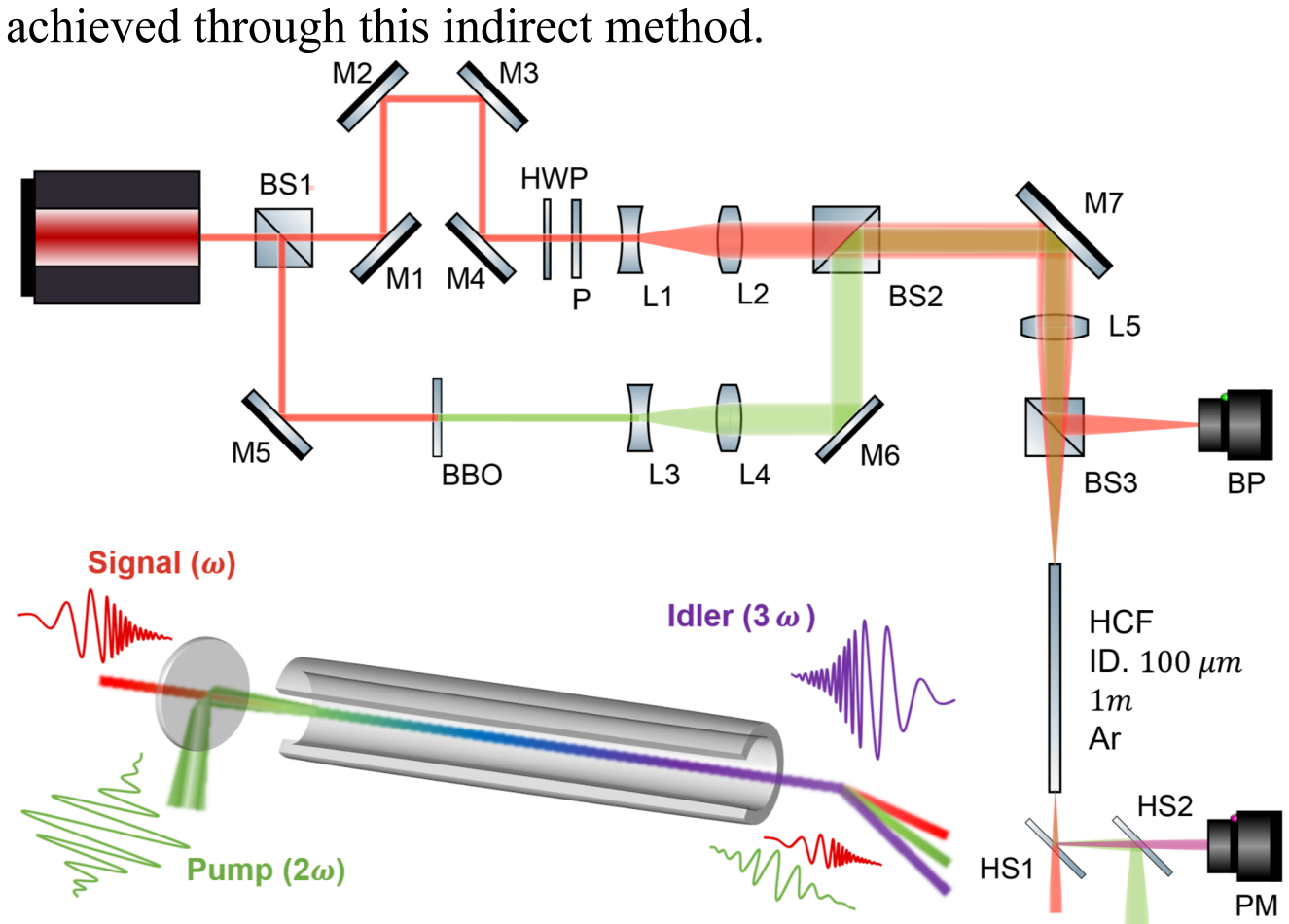

Fig.1. Schematic of the FWM experiment setup transferring pulse chirp. The input NIR beam is split into 2 arms via a beam splitter (BS1). The NIR beam goes through a delay stage (M1~M4), expanded by a telescope system ($L_1, L_2$). The other arm gets its second harmonic by a BBO and is filtered by a harmonic separator. The second harmonic pulse is expanded by a telescope ($L_3$, $L_4$) and recombined by a dichroic mirror (BS2). Both frequencies contain a pulse stretcher (not shown in the diagram), and they are focused by a lens ($L_5$, $FL\ =\ 50\ cm$) before the HCF, after being slightly split by an uncoated beam splitter (BS3), where the reflected pulse is incident on a beam profiler (BP), to record the virtual focus.

In this Letter, we demonstrate antisymmetric transfer of second-order dispersion (SOD) from NIR signal pulses to a UV idler, with a moderate 13% conversion efficiency under an SPM-managed operating point to preserve chirp fidelity. Unlike previous CFWM-based UV generation, our approach provides deterministic, programmable SOD transfer with opposite sign between NIR and UV, establishing an experimentally calibrated control law. We investigate this chirp transfer in degenerate CFWM ($\omega_i = 2\omega_p - \omega_s$) with numerical simulations and experiments.

Figure 1 sketches the CFWM setup and key theory. We begin with a model assuming an instantaneous $\chi^{(3)}$ response and an undepleted pump. When the pump and signal have comparable bandwidths, the idler spectrum is a triple convolution of the inputs, which smooths fine spectral features by mixing frequency components. In the time domain, chirped pulses map frequency to time, so the nonlinear polarization mixes co-temporal components and the instantaneous idler frequency follows $\omega_i(t) = 2\omega_p(t) - \omega_s(t)$. In the frequency domain, the generated complex electric field $\widetilde{E}_{CFWM}$ will be proportional to all the input pump and signal fields $\widetilde{E}_p^{\,2}\widetilde{E}_s^{*}$:

$$\widetilde{E}_{idler}(\omega) \propto i\iint \widetilde{E}_p(\omega')\widetilde{E}_p(\omega'')\widetilde{E}_s(\omega' + \omega'' - \omega)\ \eta(\omega', \omega'', \omega' + \omega'' - \omega)\, d\omega' d\omega'' \qquad (1)$$

where $\omega'$ and $\omega''$ both represent the two pump photon frequencies. We can obtain the spectral electric field at each frequency $\omega$ by calculating the sum of all contributions. The double integral arises from the fact that the degenerate pump photons appear twice, and the pump frequency appears twice in the convolution. Combining the two pump terms in Eq. (1) yields the same form as the idler in an optical parametric amplifier, in particular for difference-frequency generation. Because the signal enters as a complex conjugate with a negative frequency term, higher-frequency signal components map to lower-frequency idler components, and vice versa. Consequently, the idler inherits the signal's dispersion with the opposite chirp direction.

**Results.** We first simulate the antisymmetric chirp transfer in Fig. 2(a, b). In Fig. 2(a), the pump, signal, and idler spectral intensity peaks coincide at zero delay. In Fig. 2(b), introducing a $160\,fs$ temporal signal advance shifts the idler peak proportionately, consistent with the convolution model. Then, we validate the calculation results via the setup in Fig. 1. The laser system used is a 20W, 5kHz Ytterbium-doped potassium gadolinium tungstate (Yb, PHAROS-SP, Light-Conversion™) with a transform-limited pulse duration of approximately $220\,fs$. A delay line is incorporated into the signal path to ensure temporal overlap of the signal and pump pulses inside the HCF. A motorized multi-grating stretcher imposes the desired GDD on the signal, while the pump is kept transform-limited. The HCF diameter is 100 μm, and 1 m in length. To maximize the coupling efficiency into the HCF, the focal spot sizes of both frequencies are around $65um$, controlled by the telescopes. The group-velocity mismatch of the HCF is about $100\,fs/m$. Fig. 2(c-g) shows the resulting UV spectra for different signal-pump delays. Fig. 2(c) and (d) display the spectra for the NIR signal pulse with positive (GDD $\approx\ 7.9\times10^4\,fs^2$) and negative ($GDD \approx\ -8\times10^4\,fs^2$) linear chirp, respectively, with both pulses stretched to $\sim1000\,fs$. Additionally, Fig. 2(f) shows the results for a $2000\,fs$ pulse with a negative chirp ($GDD \approx\ -1.6\times10^5\,fs^2$). When the positively chirped signal pulse precedes the pump by a larger delay, the shorter wavelength components mix with the pump (right side in Fig. 1(a)), producing a UV idler shifted to longer wavelengths. As the delay decreases, the result reverses. From these spectral traces, we can deduce that a positively chirped signal induces a negatively chirped UV idler pulse. Notably, the trace map in Fig. 2(e) remains tilted at the transform-limited pump and signal with the energy of 10 μ$J$ and 5 μ$J$ before the entrance fused silica window, indicating the presence of self-phase modulation (SPM) in the HCF for the seeded pulses. SPM-induced phase accumulation perturbs the idler chirp, as seen in Fig. 2(e). By reducing the total pump and signal power to 2 μ$J$ and 1 μ$J$ per pulse, respectively, this tilt vanishes in Fig. 2(g).

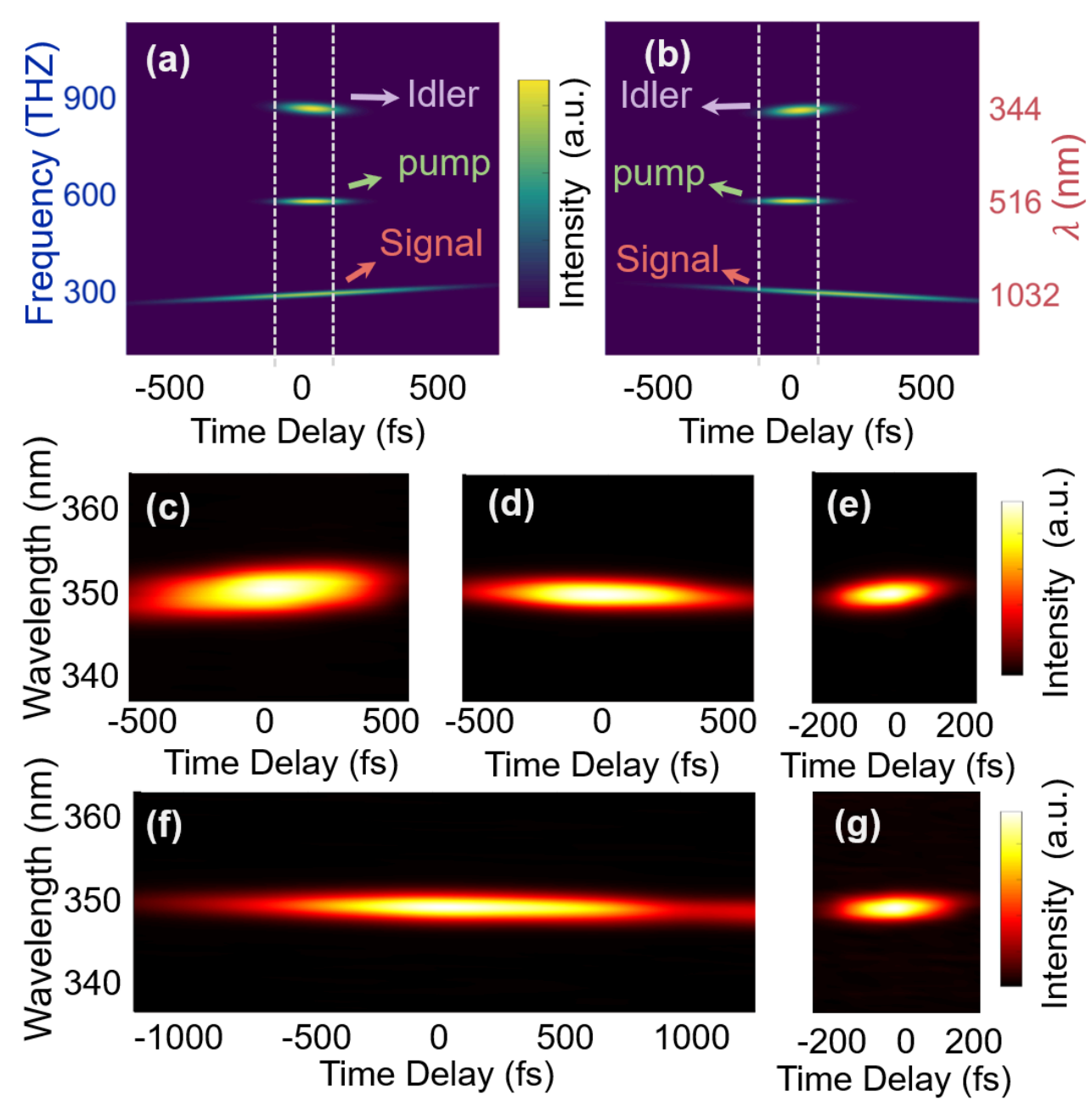

Fig. 2. Numerical calculation (a, b) and experimental results (c-g). (a, b) Numerical calculations of the CFWM spectrogram in the HCF under perfect phase-matching conditions. The signal and pump spectra are on the input side, and the generated idler spectrum is from the output side. The signal pulse in (a) and (b) has a positive and negative chirp at a delay position of 0 $fs$ and 160 $fs$. (c-g) show experimental spectral trace maps at the idler pulse for a linearly chirped signal pulse. The chirp rates were controlled by the pulse stretcher, while the pump at 516 nm is still transform-limited.

To quantify the antisymmetric chirp transfer between the signal and idler, we characterize the pulses in each frequency using a transient grating-frequency resolved optical gating (TG-FROG), and extract the amount of their GDD to compare with the theory. Under the stationary-phase approximation, and considering the comparative bandwidth of the pump and seeded signal, we obtain an approximated SOD of the generated idler, $\phi_{2i}$:

$$\phi_{2i}' \approx \phi_{2p}\phi_{2s}/(2\phi_{2s} - \phi_{2p}) \qquad (2)$$

where $\phi_2$ represents the GDD value of each frequency. Fig. 3 shows the experimental measurement results via TG-FROG. In the TG-FROG traces, the tilt of the central lobe encodes the chirp sign, and the quadratic spectral-phase curvature (red traces in Fig. 3(c,f)) flips from concave-down (negative chirp) to concave-up (positive chirp), confirming transfer from a negatively chirped signal to a positively chirped idler. The signal pulse is negatively chirped to around 1000 $fs$ (with the GDD of $\phi_{2s} \approx -8\times10^4\ fs^2$), while the pump is stretched to 1000 $fs$ to compensate for the duration difference by implementing the SOD of $\phi_{2p} \approx 6\times10^4\ fs^2$. From the retrieval in Fig. 3(f), the idler exhibits about $\phi_{2i} \approx 2.5\times10^4\ fs^2$, while the calculated value is $\phi_{2i}' \approx 2.2\times10^4\ fs^2$, given by Eq. (2). The small difference should be attributed to the SPM of the pulses in the HCF, as well as the residual retrieval error of the TG-FROG algorithm. The pulse energy in Fig. 3 is 21 μ$J$ (72 μ$J$) for the pump (signal) in the HCF, yielding ~13% pump-to-idler conversion. Increasing efficiency pushes the interaction deeper into the high-energy regime, enhancing SPM and thereby adding extra accumulated dispersion that perturbs the UV chirp. More details on the fidelity and conversion efficiency "trade-off" are discussed in another work[23]. TG-FROG retrievals were performed with the common pulse retrieval algorithm (COPRA)[24].

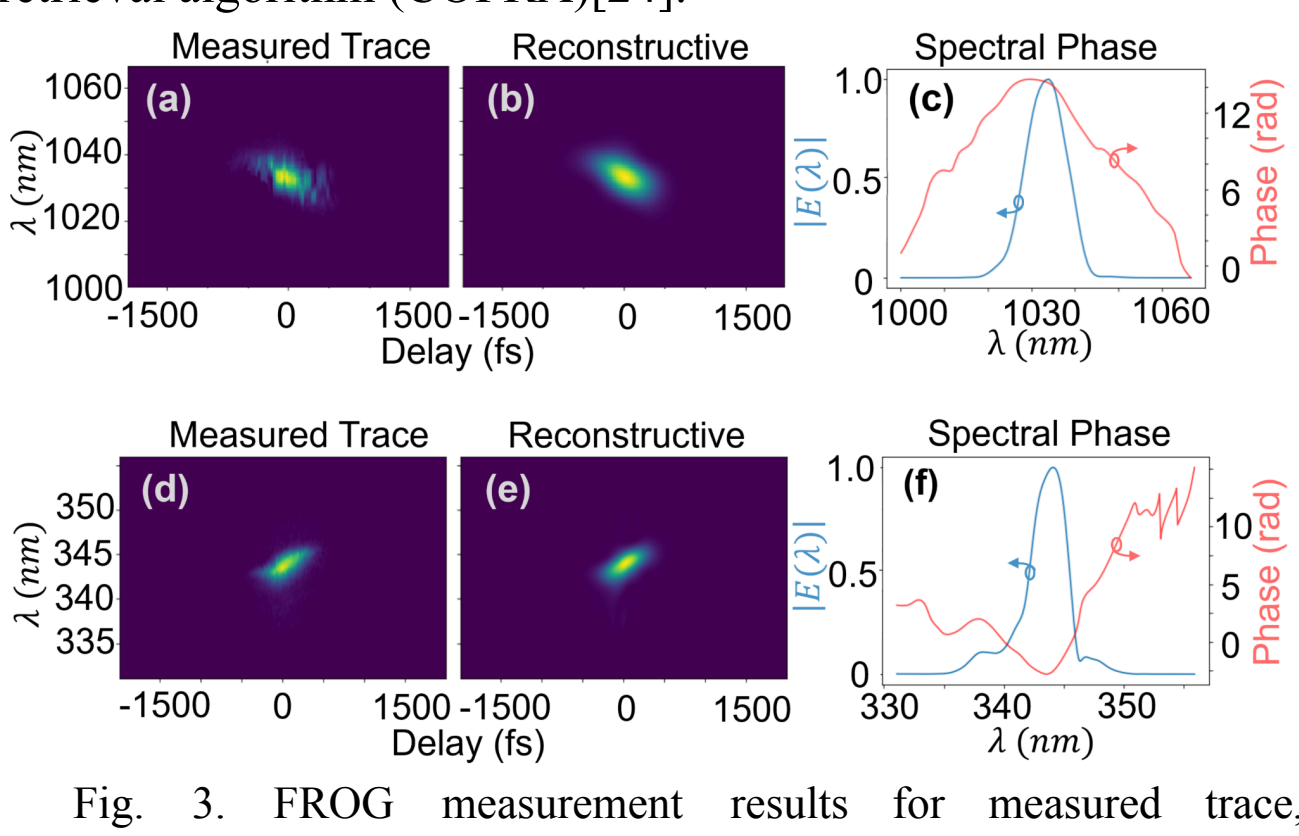

Fig. 3. FROG measurement results for measured trace, reconstructed trace, spectrum, and phase graph of the signal (a-c) and idler (d-f), respectively.

**Conversion efficiency and phase-matching conditions.** In the experiment, we balance the conversion efficiency against chirp antisymmetric transfer fidelity[23]. The signal and pump need to remain at moderate energies to suppress excessive SPM, which broadens the spectrum and contaminates the transferred phase. To keep efficiency relatively high without violating the SPM cap, we exploit the broad HCF phase-matching acceptance to improve the conversion efficiency under this moderate energy level. Moreover, the larger acceptance allows broadband frequency mixing, which can transfer the full pulse without cutting off the head and tail of the pulse spectrogram. From the nondepleted-pump approximation, we can derive the acceptance function, η, whose absolute value is given by[25]:

$$\left|\eta\left(\omega_s, \omega_p, \omega_i\right)\right| = sinc\left[\frac{1}{2}\Delta k\left(\omega_s, \omega_p, \omega_i\right)L\right] \qquad (3)$$

where $\Delta k$ is the wavevector mismatch, and $L$ represents the nonlinear interaction length. The wavevector mismatch $\Delta k$ can be expressed as

$$\Delta k = k_s + k_i - 2k_p = \Delta k_{mode} - \Delta k_{material}$$

$$= \frac{\lambda_s}{4\pi a^2}\left(u_s^2 + \frac{u_i^2}{3} - u_p^2\right) - \frac{2\pi p}{\lambda_1}\left(\delta_s + 3\delta_i - 4\delta_p\right) \qquad (4)$$

where $p$, $a$, $u_n$ are the gas pressure, core radius, and the modal constant, respectively. The gas index $n_{s,i,p} = 1 + p\delta_{s,i,p}$ is controlled by the pressure $p$. Eq. (3) describes the acceptance function of the frequency mixing process, with the wavevector mismatch, Δk, being the key parameter. Fig. 4(a, b) compares the wavevector mismatch for CFWM in HCF and SFG in the BBO crystal[15, 26], plotted versus the pump and signal wavelengths. The mismatch magnitude in the HCF case (Fig. 4(b)) is roughly three orders of magnitude smaller than in BBO (Fig. 4(a)), while the usable bandwidth is larger. Correspondingly, the sinc-acceptance of HCF is far broader, indicating that CFWM in gas-filled HCF provides a much wider phase-matching bandwidth than crystal-based SFG, which is an advantage for broadband chirp transfer without stringent angular or temperature tuning. Practically, this broad acceptance allows for choosing pressures and core sizes that keep $\Delta k$ near zero while restraining SPM by limiting pulse energies, thereby preserving the quality of the antisymmetric SOD transfer.

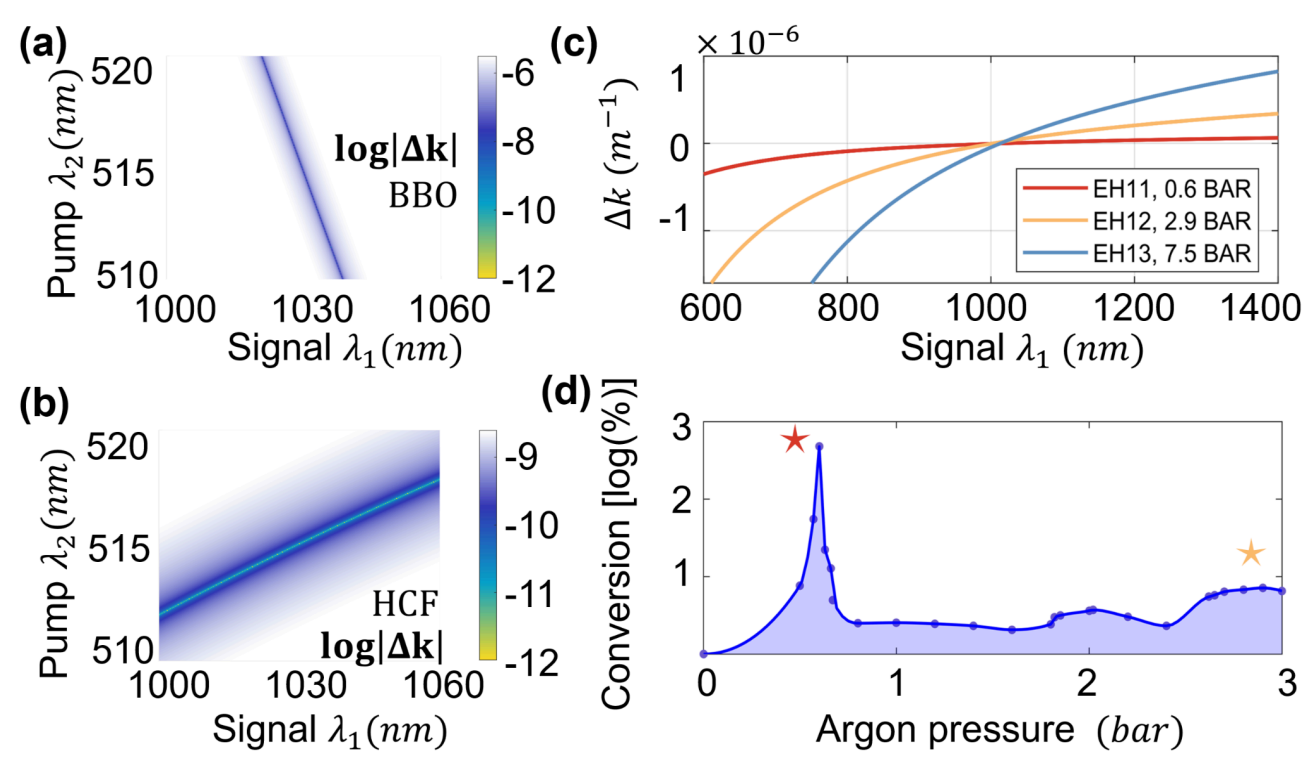

Fig. 4. (a) Phase-mismatch map for SFG in BBO, and (b) for CFWM in a gas-filled HCF, plotted versus pump and signal wavelengths; the color bar shows $log_{10}|\Delta k|$. *(c)* Phase-matching condition of Eq. (4) for different hybrid modes; red (yellow) star marks the $EH_{11}$ ($EH_{12}$) solution at 0.6 bar (2.9 bar). (d) Measured pump-to-idler conversion versus argon pressure. The intersection at 1030 nm in (c) indicates the phase-matching point used in our experiments, consistent with (d).

Nearly perfect phase matching in HCF could be achieved by balancing the normal gas dispersion with the anomalous modal dispersion from the HCF. Higher-order modes exhibit larger modal dispersion, an effect that becomes more pronounced toward higher optical frequencies. Consequently, a higher gas pressure is required to compensate for this increased modal term, as described by Eq. (4). Fig. 4(c) shows the phase-matching pressures when the idler is in the first three modes, while the signal and pump are at the lowest-order $EH_{11}$ mode. The phase-matching pressures were confirmed by the results shown in Fig. 4(d), where the red (yellow) star represents the $EH_{11}$ ($EH_{12}$) mode. Decreasing the HCF core size increases anomalous waveguide dispersion ($\propto a^{-2}$), and needs higher pressure to satisfy the phase matching condition, as shown in Eq. (4). The higher pressure can result in a larger conversion from the simple theory expression, which has shown excellent agreement with CFWM experiments in noble gas[27]:

$$I_i \propto \frac{p|\chi^3|^2 I_s I_p^2 L^2}{n_s n_p^2 \lambda_i^2} \quad (5)$$

where pressure $p$ affects the number density of the argon atoms, $\chi^3$, I, L are the third-order nonlinear coefficient, the intensity, and the interaction length, respectively. As a result, a thinner fiber under the phase-matching pressure achieves higher-energy idler generation without driving the system into an excessive SPM regime. This operating point preserves the intended antisymmetric chirp transfer from the NIR signal to the UV idler.

**Discussion.** A technical limitation in characterizing the idler via our TG-FROG arises from the minimum required energy for stable retrieval. Therefore, we operate at moderate energies that preserve fidelity while improving conversion efficiency, ensuring the idler is above the reliable detection threshold. Further improvements in pump-to-idler efficiency can be achieved by tuning the core diameter and gas species/pressure in Eq. (5), such as helium at higher phase-matching pressure. Our system can be readily adapted to high-energy UV generation with a user-defined chirp. The approach is also extendable to the deep-UV by shortening the pump wavelength and/or red-shifting the signal, and scalable to higher average power by increasing the repetition rate. Such capabilities have broad potential in high-power UV generation with user-defined programmability or machine-learning-assisted control[10, 28], attosecond physics[29, 30], laser manufacturing[31], femtosecond stimulated Raman spectroscopy[32], high-resolution spectroscopy[33], and quantum control spectroscopy[3, 34, 35].

**Conclusion.** In this Letter, we experimentally demonstrate antisymmetric transfer of SOD from NIR to UV via CFWM in a gas-filled HCF. A negative (positive) signal chirp maps to a positive (negative) UV idler with near-linear tunability, as confirmed by TG-FROG retrievals and supported by an analytical model. With pressure-tunable phase matching and moderate input energies, the scheme delivers microjoule-level UV while keeping SPM well controlled. Compared with crystal-based SFG, the gas-phase platform offers an octave-spanning phase-matching acceptance bandwidth, higher damage tolerance, and minimal walk-off, enabling broadband UV generation with high-fidelity chirp transfer. The method is readily extendable to the deep-UV by redshifting/blueshifting the signal/pump, adjusting gas species such as helium under higher phase-matching pressure[36]. The calibrated signal-to-idler chirp mapping establishes a practical foundation for forthcoming programmable UV phase shaping in HCFs, enabled by an NIR signal pulse shaper. This indirect broadband route to UV SOD tunability opens new opportunities in attosecond and ultrafast sciences, photoemission studies, and precision microfabrication.

**Funding.** The U.S. Department of Energy (DOE), the Office of Science, Office of Basic Energy Sciences, under Contract No. DE-SC0022559, No. DE-FOA-0002859, the National Science Foundation under Contract No. NSF2431903.

**Acknowledgments**. The authors thank the support from UCLA, the California NanoSystems Institute, and the SLAC National Accelerator Laboratory.

**Disclosures.** The authors declare no conflicts of interest.

**Data availability.** Data underlying the results presented in this paper are not publicly available at this time but may be obtained from the authors upon reasonable request.

**Full reference** (for review purposes only.):